\documentclass{article}

\usepackage{graphicx}
\usepackage{amsmath,amssymb,amsfonts,bm}
\usepackage[margin=1in]{geometry}
\usepackage[margin=1in]{geometry}
\usepackage{url}
\usepackage{hyperref}
\hypersetup{colorlinks=false,pdfborder={0 0 0}}

\usepackage[utf8]{inputenc}
\usepackage[english]{babel}

\usepackage{siunitx}

\newcommand{\dd}[2]{\frac{\text{d}{#1}}{\text{d}{#2}}}

\newcommand{\eps}{\epsilon}
\newcommand{\mat}[1]{\text{\bf #1}}
\newcommand{\Cov}{{\rm Cov}}

\def\T{^{\rm T}}

\begin{document}
\title{A comparison of nonlinear extensions to the ensemble Kalman filter\\Gaussian Anamorphosis and Two-Step Ensemble Filters}


\author{Ian Grooms}


\date{Department of Applied Mathematics, University of Colorado, Boulder, CO, USA }

\maketitle

\begin{abstract}
Ensemble Kalman filters are based on a Gaussian assumption, which can limit their performance in some non-Gaussian settings.
This paper reviews two nonlinear, non-Gaussian extensions of the Ensemble Kalman Filter: Gaussian anamorphosis (GA) methods and two-step updates, of which the rank histogram filter (RHF) is a prototypical example.
GA-EnKF methods apply univariate transforms to the state and observation variables to make their distribution more Gaussian before applying an EnKF.
The two-step methods use a scalar Bayesian update for the first step, followed by linear regression for the second step.
The connection of the two-step framework to the full Bayesian problem is made, which opens the door to more advanced two-step methods in the full Bayesian setting.
A new method for the first part of the two-step framework is proposed, with a similar form to the RHF but a different motivation, called the `improved RHF' (iRHF).
A suite of experiments with the Lorenz-`96 model demonstrate situations where the GA-EnKF methods are similar to EnKF, and where they outperform EnKF.
The experiments also strongly support the accuracy of the RHF and iRHF filters for nonlinear and non-Gaussian observations; these methods uniformly beat the EnKF and GA-EnKF methods in the experiments reported here.
The new iRHF method is only more accurate than RHF at small ensemble sizes in the experiments reported here. 
\end{abstract}

\section{Introduction}
\label{sec:Intro}
Data assimilation (DA) is the process of combining observational data and a dynamical model to estimate and predict the state of a dynamical system.
It encompasses a variety of methods with varying degrees of complexity, efficiency, and robustness \cite{Kalnay02,Evensen09,LSZ15}.
DA is widely used in a variety of applications including weather forecasting \cite{IFS}, atmospheric and oceanic sciences \cite{20CR}, oil \& gas reservoir modeling \cite{ANORV09,FMS20}, and epidemiology \cite{Sesterhenn20}, to name just a few.
Bayesian estimation theory provides a mathematical framework for DA (e.g.~\cite{WB07}).
Bayes' theorem relates the posterior probability distribution describing the state of the system given observations to the probability of the observations given the system state (the `likelihood') and the prior probability of the state of the system without knowing the observations.
The Bayesian \emph{filtering} problem is to find the posterior on the current system state given all past and current observations; by contrast, \emph{smoothing} problems allows future observations to impact the posterior.
The focus here is on large-scale applications as occur in the geosciences.
For example, a global ocean model with modest resolution and no biogeochemistry has in the vicinity of 20 million degrees of freedom, while coupled climate models at higher resolution can have on the order of billions of degrees of freedom.\\

When the dynamics and the observations are linear and all distributions are Gaussian, the Bayesian filtering problem is exactly solved by the Kalman filter \cite{Jaz70}.
Ensemble Kalman filtering, which comprises an entire class of methods, is an approach to solving the filtering problem based on ensemble approximations to the Kalman filter formulas \cite{Evensen09}.
It is known that the mean and covariance generated by ensemble Kalman filters (EnKFs) converge to those of the Kalman filter in the limit of infinite ensemble size for linear systems with Gaussian noise \cite{MCB11,KM15}.
EnKFs are routinely used in systems with nonlinear dynamics and non-Gaussian distributions but there is no guarantee of convergence for these systems.
Indeed, EnKFs are expected {\it not} to converge to the correct Bayesian posterior for nonlinear, non-Gaussian systems.
Nevertheless, EnKFs are relatively easy to implement compared to alternatives like 4D-Var \cite{Talagrand10} and yield surprisingly good results in many nonlinear problems.

There is a group of methods that aims to extend EnKF to improve performance in problems with non-Gaussian distributions and nonlinear observations while retaining algorithmic similarity to EnKFs.
This group includes Gaussian mixture methods \cite{AA99,BSN03,HPTK08,SKNSV11,FK13b,LAH16,AMS18,NRCB18}, methods based on\\
gamma/inverse-gamma distributions \cite{Bishop16,PB18}, methods that target higher moments of the posterior \cite{Hodyss12,HACCR17}, methods based on rank statistics \cite{Anderson10,MCSB14,Anderson19,Anderson20}, and `Gaussian Anamorphosis' methods \cite{BEW03,BPW10,ZGFL11,BrankartEtAl12,SB12}.
Gaussian anamorphosis (GA) methods were originally motivated by the desire to keep certain state variables - like concentration, mass, or volume - positive, a constraint not respected by EnKFs.
GA methods operate by applying changes of variable to the prior to make it closer to Gaussian, then applying the EnKF, then transforming back.
The rank-histogram filter (RHF) methods of \cite{Anderson10,MCSB14,Anderson19,Anderson20} are another example of methods that aim to relax Gaussian assumptions while retaining algorithmic similarity to EnKFs; the RHF and related methods operate within a two-step assimilation framework developed by Anderson \cite{Anderson03} that unites Bayesian estimation and linear regression.

The focus here is on nonlinear and non-Gaussian extensions of ensemble Kalman filters.
The two main categories of competing methods, in the context of large-scale geoscience applications, are variational methods and particle filters.
Though these are outside the scope of the review, a brief overview is included here for interested readers.

Variational methods \cite{Talagrand10} construct an objective function - usually proportional to the negative log of the posterior - with two terms: one that penalizes deviations from the prior mean and one that penalizes deviations from observations.
Optimization methods are then used to seek a global minimizer, which is the Maximum A Posteriori (MAP) estimator of the true state, though in practice the methods might only find a local minimizer.
Variational methods can be crudely divided into two broad categories as follows: In 4D-Var there are observations of the state at future times, whereas in 3D-Var there are not.
4D-Var handles nonlinearity in both the dynamics and observations more accurately than basic EnKF-type methods since it computes the MAP based on the true posterior.
To reduce cost, 4D-Var is often applied to a sequence of shorter time windows instead of to a single long time window.
One main drawback of classical 4D-Var is that the penalty on the initial condition does not vary from one time window to the next; for example, the forecast error model is the same on a rainy day as on a dry one.
The ability of 4D-Var to provide excellent state estimates has resulted in widespread use in operational weather and ocean DA \cite{FischerEtAl05,RawlinsEtAl07,HuangEtAl09,MooreEtAl11,GCHRO13}.
One way to improve the performance of 4D-Var is to use an ensemble forecast to codify the forecast uncertainty, which can then be used to define the term penalizing deviations from the forecast mean in the 4D-Var objective function.
There are many ways to do this \cite{Bannister17}; in `pure' ensemble-variational (EnVar) methods the ensemble forecast covariance matrix is used directly, while in `hybrid' EnVar the ensemble forecast covariance is combined with a time-independent background.
These EnVar methods are the current state of the art for numerical weather prediction \cite{WL14,IFS}.

Particle filters are a class of sequential Monte Carlo methods \cite{DdFG01} that solve the Bayesian smoothing and filtering problems via sequential importance sampling.
The classical bootstrap or `SIR' particle filter of \cite{GSS93} generates an empirical measure that provably converges weakly to the correct posterior distribution in the limit of an infinite ensemble size \cite{CD02,LSZ15}, but the convergence is prohibitively slow in high-dimensional systems \cite{BBL08,SBBA08,SBM15}.
In view of this limitation there is a large body of research aiming to design alternatives to the classical particle filter for applications with extremely high dimension, including \cite{CT09,CMT10,CT12,vanLeeuwen09,AvL13,AvL15,PM15,RvH15,Poterjoy16,RK17,RGK18,PWH19,PvL19,PvLG19,HvL21}.
Particle filters have also been hybridized with EnKFs \cite{FK13a,CRR16,RG20b} and variational methods \cite{MHP18}.

The goal of the present paper is to compare GA methods and the RHF as nonlinear extensions of the EnKF.
We begin with a review of GA methods and of the two-step framework on which both the GIGG-EnKF (Gamma, Inverse-Gamma, and Gaussian; \cite{Bishop16,PB18}) and the RHF are built.
New versions of both the GA and RHF filters are presented.
Then, GA and RHF methods are tested in the Lorenz-`96 model with nonlinear observations, with an EnKF for comparison.
In these tests the RHF and its newly-proposed extension are the most accurate, and do not incur significant extra computational cost.

\section{Gaussian Anamorphosis in the Ensemble Kalman Filter}
\subsection{Ensemble Kalman Filters}\label{sec:EnKF}
Let the vector of variables that we are trying to estimate be denoted by $\bm{x}$, and let our prior uncertainty about $\bm{x}$ be encoded in the distribution of a random vector $\bm{X}$.
Assume that there is another random vector $\bm{Y}$ jointly distributed with $\bm{X}$, and we have a single `observation' vector $\bm{y}$ which is a realization of $\bm{Y}$.
For example, one might have $\bm{Y} = \bm{H}(\bm{X}) + \bm{\eps}$ where $\bm{\eps}$ is a random variable independent of $\bm{X}$ that represents observation error, and $\bm{H}$ some nonlinear function.
This type of relationship between $\bm{X}$ and $\bm{Y}$ is widely assumed to hold in the DA literature, but is not by any means required, and will not be used in some of the numerical experiments reported below.
The prior is the marginal distribution on $\bm{X}$, the likelihood is the probability density function (pdf) of $\bm{Y}|\bm{X}$ evaluated at $\bm{Y}=\bm{y}$ and then viewed as a function of the values of $\bm{X}=\bm{x}$, and the posterior distribution that we seek is the conditional distribution $\bm{X}|\bm{Y}=\bm{y}$.

The usual derivation of the Kalman Filter (KF) formulas that form the foundation for the EnKF assumes that $\bm{X}$ has a Gaussian distribution with mean $\bm{x}_b$ and covariance $\mat{B}$, that $\bm{Y} = \mat{H}\bm{X}+\bm{\eps}$ where $\mat{H}$ is a constant matrix (independent of $\bm{X}$ and $\bm{\eps}$) and $\bm{\eps}$ is a centered Gaussian random vector independent of $\bm{X}$ with covariance $\mat{R}$.
An equivalent and particularly useful alternative derivation of the KF formulas makes a Gaussian approximation of the joint distribution of $\bm{X}$ and $\bm{Y}$.
It is well known that if a joint distribution is Gaussian then the conditional is also Gaussian, with mean and covariance
\begin{equation}\label{eqn:GaussCondMean}
\mathbb{E}[\bm{X}|\bm{Y}=\bm{y}] = \bm{x}_b + \Cov[\bm{X},\bm{Y}]\Cov[\bm{Y}]^{-1}(\bm{y}-\bar{\bm{y}})
\end{equation}
and
\begin{multline}\label{eqn:GaussCondCov}
\Cov[\bm{X}|\bm{Y}=\bm{y}] = \\\Cov[\bm{X}] - \Cov[\bm{X},\bm{Y}]\Cov[\bm{Y}]^{-1}\Cov[\bm{Y},\bm{X}].
\end{multline}
Under the assumptions stated above we find
\begin{eqnarray}
\bar{\bm{y}} &=& \mat{H}\bm{x}_b,\\
\Cov[\bm{X}] &=& \mat{B},\\
\Cov[\bm{X},\bm{Y}] &=& \mat{BH}\T,\\
\Cov[\bm{Y}] &=& \mat{HBH}\T+\mat{R}.
\end{eqnarray}
Inserting these into equations (\ref{eqn:GaussCondMean}) and (\ref{eqn:GaussCondCov}) yields the familiar Kalman update formulas
\begin{eqnarray}\label{eqn:KF_Mean}
\bm{x}_a&=&\bm{x}_b + \mat{K}(\bm{y}-\mat{H}\bm{x}_b),\\\label{eqn:KF_Cov}
\mat{C}&=&\left(\mat{I}-\mat{KH}\right)\mat{B},\\\label{eqn:KF_Gain}
\mat{K}&=&\mat{BH}\T\left(\mat{HBH}\T+\mat{R}\right)^{-1},
\end{eqnarray}
where $\bm{x}_a$ is the posterior mean, also called the analysis mean, $\mat{C}$ is the posterior covariance, and $\mat{K}$ is the Kalman gain matrix.

The first EnKF was proposed by Evensen \cite{Evensen94}.
An equally-weighted ensemble was used to obtain an approximation of the background covariance matrix $\mat{B}$, and then the Kalman update formula (\ref{eqn:KF_Mean}) with $\bm{x}_b$ replaced by $\bm{x}_i$ was applied to each ensemble member, where $\bm{x}_i$ with $i=1,\ldots,N$ denotes an ensemble of $N$ state vectors.
The original EnKF contained a flaw in that while the analysis ensemble mean was consistent with (\ref{eqn:KF_Mean}), the analysis ensemble covariance was not consistent with (\ref{eqn:KF_Cov}).
This was simultaneously fixed in \cite{BLE98,HM98} using the update formula
\begin{equation}
\bm{x}_i^+=\bm{x}_i + \mat{K}(\bm{y}-\mat{H}\bm{x}_i-\bm{\eps}_i),
\end{equation}
where $\bm{\eps}_i$, $i=1,\ldots,N$ are independent samples from the distribution of $\bm{\eps}$, and the superscript $^+$ denotes the analysis ensemble members.
This version of the EnKF has come to be known as `the' EnKF, or the `stochastic' EnKF, or the `perturbed-observations' EnKF.
Several alternative EnKFs that do not require an ensemble of observation errors were proposed in later years \cite{PVG98,Anderson01,BEM01,WH02}; these can be called deterministic EnKFs.
See \cite{HZ16} for a review of the literature on EnKFs for atmospheric data assimilation.


\subsection{Gaussian Anamorphosis}
\label{sec:GA}
As noted above, the EnKF uses a Gaussian approximation of the joint distribution of $\bm{X}$ and $\bm{Y}$, which is patently erroneous when the scalar marginal distributions, i.e. the distributions of each element of $\bm{X}$ and $\bm{Y}$, are far from normal.
In such cases the Gaussian approximation can be relaxed by transforming to variables with Gaussian marginals, then applying the EnKF, then transforming back.
The idea of using univariate transformations of state variables in the EnKF was first proposed in \cite{BEW03}, with the main goal being to force the EnKF update to respect bounds on the variables, e.g. to prevent the EnKF from producing negative mass.
There are other methods for dealing with linear inequality constraints in the EnKF, e.g. by using truncated normals \cite{LBCBBV09} or using constrained optimization \cite{JMCV14,ABLSS19,LJHP19}. The point here is not to deal with inequality constraints per se, but to introduce transforms as a means of generalizing the EnKF to non-Gaussian distributions.

Some methods within the literature explicitly aim to construct the univariate transformations so that the marginals are all Gaussian, while others merely design their transformations so that the state variables are guaranteed to respect desired boundedness properties like positivity after the EnKF update.
The former are known as `Gaussian anamorphosis' methods \cite{BEW03} and `normal score' methods \cite{ZGFL11,LZHG12} in the ensemble DA literature.
The correct probability-integral transformation of a real-valued scalar random variable $X$ to a standard normal random variable $\hat{X}$ is $\hat{X} = \Phi^{-1}(F_X(X))$ where $F_X$ is the cumulative distribution function (cdf) of $X$ and $\Phi$ is the standard normal cdf.
This only applies when $X$ has a non-atomic density; it is not possible to transform $X$ to a standard normal when the $F_X$ is discontinuous.
The ad hoc transformations appearing in the literature involve poor (but still consistent) estimators of $F_X$ using samples of $X$, mainly based around piecewise-linear  non-parametric approximations (e.g. \cite{SB12}).
Application of these methods has been fairly limited, comprising ocean ecosystem models \cite{BBBOV10,SB12,DBBLM13,GSBSKD17}, groundwater models \cite{ZGFL11,LZHG12}, and some limited use in meteorological applications \cite{LKM13,LKM16a,LKM16b,KMTLK17}.

Using GA transformations in the EnKF relaxes the EnKF approximation that the joint distribution of $\bm{X}$ and $\bm{Y}$ is Gaussian to an approximation that the copula is Gaussian (see the appendix  for a definition of the copula of jointly-distributed random variables).
As emphasized in \cite{AvL14}, this updated approximation is not necessarily an improvement, and can in fact lead to {\it worse} results than a standard EnKF.
On the other hand, practical experience \cite{BBBOV10,ZGFL11,LZHG12,SB12,DBBLM13,LKM13,LKM16a,LKM16b,KMTLK17,GSBSKD17} and many of the examples in \cite{AvL14} show that GA usually improves the performance significantly.
Furthermore, GA is a legitimate non-Gaussian extension of the EnKF: GA methods should converge towards sampling from the true posterior for a wider range of problems than the EnKF, namely problems with Gaussian copulas.
A proof of this conjecture is beyond the scope of the current article.
Multivariate transformations based on optimal transport \cite{MMPS16} could potentially be used to relax even the assumption of a Gaussian copula, and there are some promising preliminary results using these transformations in the context of data assimilation \cite{SBM19}.

We close this review of GA methods by proposing a perturbed-observations GA-EnKF:
\begin{enumerate}
\item Start with an ensemble $\bm{x}_i$, $i=1,\ldots,N$ and use it to generate an ensemble $\bm{y}_i$, as in the stochastic EnKF, by generating $\bm{y}_i$ as a realization of $\bm{Y}|\bm{X}=\bm{x}_i$.
\item Use the ensembles to find approximations to the marginal cdfs of the elements of $\bm{X}$ and $\bm{Y}$.
\item Apply univariate transformations of the form $\Phi^{-1}(F(\cdot))$ to each scalar state variable in the ensembles (a different transform for each variable) to get transformed ensembles $\hat{\bm{x}}_i$, $\hat{\bm{y}}_i$ and a transformed observation vector $\hat{\bm{y}}$.
\item Apply the perturbed-observation EnKF in the transformed space using the formula
\begin{equation}
\hat{\bm{x}}_i^+=\hat{\bm{x}}_i + \Cov[\hat{\bm{X}},\hat{\bm{Y}}]\Cov[\hat{\bm{Y}}]^{-1}(\hat{\bm{y}}-\hat{\bm{y}}_i).
\end{equation}
In this update step, use ensemble approximations to $\Cov[\hat{\bm{X}},\hat{\bm{Y}}]$ and $\Cov[\hat{\bm{Y}}]$.
\item Apply inverse transformations of the form $F^{-1}(\Phi(\cdot))$ to the updated ensemble members.
\end{enumerate}
A simple yet crucial aspect of the proposed algorithm is the use of the conditional Gaussian formula (\ref{eqn:GaussCondMean}) rather than the KF formulas (\ref{eqn:KF_Mean}) and (\ref{eqn:KF_Gain}).
There has been some confusion in the literature surrounding the question of how to use the KF update formulas in the transformed space, since the KF formulas need $\mat{H}$ and $\mat{R}$, which are not clearly defined in the transformed space.
Note that the above GA-EnKF algorithm does not specify details for estimating the marginal cdfs $F$; two specific methods for approximating the cdfs $F$ and their inverses are detailed in section \ref{sec:GA_Details}.

Step 4 uses ensemble approximations of $\Cov[\hat{\bm{X}},\hat{\bm{Y}}]$ and $\Cov[\hat{\bm{Y}}]$.
There is, of course, no need to form these matrices, which are too big to fit in memory for many applications; their factored form in terms of scaled ensemble perturbation matrices can be stored and used instead.
Let $\hat{\mat{A}}_X$ be a matrix whose columns are
\begin{equation}
\frac{1}{\sqrt{N-1}}\left(\hat{\bm{x}}_i-\hat{\bar{\bm{x}}}\right)
\end{equation}
and similary for $\hat{\mat{A}}_Y$.
Then the update formula becomes
\begin{equation}
\hat{\bm{x}}_i^+=\hat{\bm{x}}_i + \hat{\mat{A}}_X\hat{\mat{A}}_Y\T\left(\hat{\mat{A}}_Y\hat{\mat{A}}_Y\T\right)^{-1}(\hat{\bm{y}}-\hat{\bm{y}}_i).
\end{equation}
If the observation errors are independent, serial scalar assimilation can be used in the transformed space.
If simultaneous assimilation is desired or required and the ensemble size $N$ is small enough that $\hat{\mat{A}}_Y\hat{\mat{A}}_Y\T$ is not invertible, then Schur-product localization (or some other regularization, e.g. \cite{Pourahmadi11}) can be used to generate an invertible approximation to $\Cov[\hat{\bm{Y}}]$.

Note that the transformations used in GA are univariate, and thus do not change the spatial locations of the variables, i.e. the spatial location of the $j^\text{th}$ element of $\bm{x}$ is the same as the spatial location of the $j^\text{th}$ element of $\hat{\bm{x}}$.
It is thus reasonable to suppose that the optimal length scales for localization in the transformed space may be similar to those in the original space.
Similarly, the fact that the transformations are univariate implies that observation errors that are independent before transformation remain independent after transformation, which is important if one intends to use serial assimilation in the transformed space.

Deterministic EnKFs can also be designed based on the conditional Gaussian formulas (\ref{eqn:GaussCondMean}) and (\ref{eqn:GaussCondCov}).
For example, an ETKF update \cite{BEM01,WBJ04} of the scaled transformed ensemble perturbation matrix $\hat{\mat{A}}_X$ could be accomplished as follows.
To exactly satisfy (\ref{eqn:GaussCondCov}), the updated scaled transformed ensemble perturbation matrix $\hat{\mat{A}}_{X,+}$ should satisfy
\begin{equation}
\hat{\mat{A}}_{X,+}\hat{\mat{A}}_{X,+}\T = \hat{\mat{A}}_X\left(\mat{I}-\hat{\mat{A}}_Y\T\left(\hat{\mat{A}}_Y\hat{\mat{A}}_Y\T\right)^{-1}\hat{\mat{A}}_Y\right)\hat{\mat{A}}_X\T.
\end{equation}
If $\mat{T}$ is a symmetric square root 
\begin{equation}
\mat{T} = \left(\mat{I}-\hat{\mat{A}}_Y\T\left(\hat{\mat{A}}_Y\hat{\mat{A}}_Y\T\right)^{-1}\hat{\mat{A}}_Y\right)^{1/2},
\end{equation}
then 
\begin{equation}
\hat{\mat{A}}_{X,+} = \hat{\mat{A}}_X\mat{T}
\end{equation}
will exactly satisfy (\ref{eqn:GaussCondCov}).
This ETKF update in the transformed space could be localized using either a local analysis \cite{BrusdalEtAl03,HKS07} or the gain form of the ETKF \cite{BWL17}.

\section{Two-Step Bayesian Updating}
In a much-cited paper \cite{Anderson03}, Anderson developed a widely-used two-step framework for ensemble DA based on a combination of Bayesian estimation and least-squares regression.
The connection of the two-step framework to the full Bayesian estimation problem is developed here.
The two-step framework of \cite{Anderson03} is based on serial assimilation of independent scalar observations (though it can be extended to batches of observations), and thus deals with scalar observations $y$.
It is straightforward to generalize to vector observations, but the presentation here is limited to the scalar case for clarity.

First, introduce a scalar random variable $Z$ with the property that $Y$ conditioned on $Z$ is independent of $\bm{X}$.
In the traditional case where $Y=H(\bm{X})+\eps$ one could let $Z=H(\bm{X})$.
If the observation $Y$ is a function of a single element of $\bm{X}$, then one could let $Z$ be the observed element of $\bm{X}$.
A general approach to choosing $Z$ is discussed in section \ref{sec:ListOfMethods}.

Next note that Bayes' rule for this problem can be written as follows
\begin{eqnarray}\notag
[\bm{x}|y] &=& \int[\bm{x},z|y]\text{d}z\\\notag
 &\propto&\int[y|\bm{x},z][\bm{x},z]\text{d}z \\\notag
 &=& \int\left([y|z][z]\right)[\bm{x}|z]\text{d}z \\
 &\propto&\int[z|y][\bm{x}|z]\text{d}z
\end{eqnarray}
where the notation $[\cdot]$ represents the pdf of a random variable and the proportionality constants are required to normalize the expressions into pdfs.
Sampling from the posterior can thus be performed in a two-step process: First draw samples $z_i^+$, $i=1,\ldots,N$ from the posterior distribution $Z|Y=y$, then draw $\bm{x}_i^+$ from the conditional distribution $\bm{X}|Z=z_i^+$.
No assumptions of Gaussianity have been made in deriving this two-step update.

The first step involves sampling from a Bayesian posterior on a scalar random variable, which can be accomplished by any of a variety of methods, including the ensemble adjustment Kalman filter (EAKF; \cite{Anderson01,Grooms20}) as originally proposed by Anderson \cite{Anderson03}.
Any one of a variety of alternatives could be used, including particle filters, Gaussian mixture models, and GA-EnKFs,  though no publications have developed these options in the two-step filtering context.
The GIGG-EnKF \cite{Bishop16,PB18} and the rank histogram filter (RHF; \cite{Anderson10}) are specifically developed as options for the first step of the two-step framework.
These are all examples of methods that use deterministic transformations to transform a sample from the prior on $Z$ to a sample from the posterior $Z|Y=y$.
The two-step framework in principle also enables the assimilation of categorical observations, i.e. situations where the range of $Y$ is discrete, as long as a method can be constructed for sampling from the scalar posterior $Z|Y=y$.

The second step in the two-step framework was performed using linear regression by \cite{Anderson03}.
To understand the connection, note that the linear model
\begin{equation}
\bm{X} = \bm{\beta}_0 + \bm{\beta}_1Z+\bm{\eta}
\end{equation}
where $\bm{\eta}$ is a centered Gaussian random vector independent of $Z$ and with covariance $\bm{\Sigma}$ implies that $\bm{X}|Z=z\sim\mathcal{N}(\bm{\beta}_0 + \bm{\beta}_1z,\bm{\Sigma})$.
Anderson \cite{Anderson03} finds estimates $\hat{\bm{\beta}}_0$ and $\hat{\bm{\beta}}_1$ of the regression coefficients from the prior ensemble $\bm{x}_i,\bm{z}_i$ using ordinary least squares (the hat serves to distinguish the true coefficients from their estimates), and then computes the residuals
\begin{equation}
\bm{\eta}_i = \bm{x}_i-\hat{\bm{\beta}}_0 - \hat{\bm{\beta}}_1z_i.
\end{equation}
Approximate samples from the posterior (the distribution is not exactly the posterior unless $\hat{\bm{\beta}}_j=\bm{\beta}_j$) are generated as
\begin{equation}
\bm{x}_i^+ = \hat{\bm{\beta}}_0 + \hat{\bm{\beta}}_1z_i^++\bm{\eta}_i.
\end{equation}
Re-use of the residuals in this manner avoids the need to estimate the covariance $\bm{\Sigma}$, and essentially amounts to a deterministic method of transforming an ensemble sampled from $\bm{X}|Z=z_i$ to samples from $\bm{X}|Z = z_i^+$.

The above discussion clarifies the relationship between regression in the second step of the two-step framework of \cite{Anderson03} and the full Bayesian update, and shows the way to enabling EnKF-like algorithms for a wide range of distributions.
For example one might use a general linear model of the form
\begin{equation}
\bm{X} = \bm{\eta} + \sum_{j=1}^J\bm{\beta}_j\phi_j(z) = \bm{\eta}+\bm{\mu}(z)
\end{equation}
where the coefficients $\bm{\beta}_j$ are estimated using ordinary least squares.
Polynomial regression is a simple example where $\phi_j(z) = z^j$, though polynomial regression is likely to prove unstable whenever any of the $z_i^+$ lies outside the range of the prior ensemble $z_i$.
General linear models correspond to a conditional distribution of the form $\bm{X}|Z=z\sim\mathcal{N}(\bm{\mu}(z),\bm{\Sigma})$.
The model could be updated by incorporating heteroscedasticity, i.e. by allowing the variance of $\bm{\eta}$ to be a function of $z$, which would update the conditional distribution to $\bm{X}|Z=z\sim\mathcal{N}(\bm{\mu}(z),\bm{\Sigma}(z))$.
In a model incorporating heteroscedasticity, the residuals $\bm{\eta}_i$ would not be simply re-used in the update, but would instead require element-wise transformations of $\bm{\eta}_i$.
A range of non-Gaussian conditional distributions could be accommodated by generalized linear models of the form
\begin{equation}
\bm{g}(\bm{X}) = \bm{\eta} + \sum_{j=1}^J\bm{\beta}_j\phi_j(z)
\end{equation}
where $\bm{g}$ is a vector link function, which must be invertible.
The link function could be used, for example, to maintain bounds on the state variables.
Extensions to heteroscedasticity and generalized linear models would require techniques more sophisticated than ordinary least squares to estimate the coefficients $\bm{\beta}_j$, but the overall two-step framework remains the same:
\begin{enumerate}
\item Estimate the unknown model coefficients $\bm{\beta}_j$ using the data in the prior ensemble
\item Obtain samples $z_i^+$ from $Z|Y=y$
\item Obtain samples from the posterior using
\begin{equation}
\bm{x}_i^+ = \bm{g}^{-1}\left(\bm{\eta}_i^++ \sum_{j=1}^J\bm{\beta}_j\phi_j(z_i^+)\right)
\end{equation}
where, for a homoscedastic model, $\bm{\eta}_i^+=\bm{\eta}_i$.
\end{enumerate}
The methods of \cite{MCSB14,Anderson19,Anderson20} are aimed to replace linear regression in the second step with nonlinear transformations.
Novel models for the second step of the two-step process are not explored further here.

\section{An Improved Rank Histogram Filter}
This section reviews a specific algorithm for the first step of the two-step process above, namely the Rank Histogram Filter (RHF; \cite{Anderson10}), and then proposes a similar algorithm that is expected to be more accurate at small ensemble sizes.

\subsection{The Rank Histogram Filter}
The RHF \cite{Anderson10} transforms samples from the prior on $Z$ into samples from the posterior $Z|Y=y$ using an approximation of
\begin{equation}
z_i^+ = F_+^{-1}(F_Z(z_i))
\end{equation}
where $F_Z$ is the prior cdf of $Z$ and $F_+$ is the posterior cdf of $Z|Y=y$.
Both cdfs are estimated from piecewise approximations of the prior and the likelihood.
Let $z_{k_1}<\ldots<z_{k_N}$ be a sorted permutation of the samples $z_1,\ldots,z_N$.
(Note that for non-atomic distributions the samples $z_i$ can be sorted without ties with probability 1.)
To represent the prior, the range of $Z$ is broken into $N+1$ subintervals where the left and right boundaries are the boundaries of the domain, and the interior boundaries are $z_{k_i}$.
Each subinterval is assumed to account for total probability $1/(N+1)$.
The prior pdf is represented as piecewise-constant on each of the interior subintervals:
\begin{equation}
p(z) = \frac{z_{k_{i+1}}-z_{k_i}}{N+1}\text{ for }z\in[z_{k_i},z_{k_{i+1}}).
\end{equation}
In the tails $p(z)$ is represented as a Gaussian pdf using the empirical mean and standard deviation of the sample, and with each tail rescaled to have total probability $1/(N+1)$.
(When the domain of $Z$ is not $\mathbb{R}$ one can instead set one or both tails to piecewise-constant, as needed.)
From this representation it is straightforward to obtain the prior cdf $F_Z$; it is only necessary to evaluate this cdf at $z_{k_i}$, where it takes values $i/(N+1)$.

The RHF approximates the likelihood $[y|z]$ as a piecewise function defined on the same subintervals as the prior.
On the interior subintervals the likelihood is simply the piecewise-linear interpolant of the likelihood evaluated at $z_{k_i}$.
In the original presentation of \cite{Anderson10} the likelihood was modeled as Gaussian in the tails, though the simpler form of approximating it as a constant in the tails was adopted more recently by Anderson in \cite{Anderson19}, and this simpler form is used here.
Specifically
\begin{equation}
\ell(z) = \left\{\begin{array}{cl}
[y|z_{k_1}] & \text{ for }z\le z_{k_1},\text{ and}\\
\left[y|z_{k_N}\right] &\text{ for } z_{k_N}\le z.
\end{array}\right.
\end{equation}
Note that the notation $\ell(z)$ is used here to denote the approximation to the likelihood, rather than the likelihood itself, which is simply denoted $[y|z]$.
With these approximations to the prior and likelihood, the approximation to the posterior becomes a piecewise function with Gaussian tails and a piecewise-linear (but discontinuous) interior:
\begin{equation}
[z|y]\approx \frac{1}{P_0}p(z)\ell(z)
\end{equation}
where 
\begin{equation}
P_0 = \int p(z)\ell(z)\mathrm{d}z
\end{equation}
and the integral is over the range of $Z$.
The integral is available in closed form.
The RHF approximation to the posterior cdf is
\begin{equation}
F_+(z)\approx \frac{1}{P_0}\int_{-\infty}^zp(s)\ell(s)\mathrm{d}s
\end{equation}
where $s$ is a dummy integration variable.
This approximation is an error function in the tails, and piecewise-quadratic in the interior, and can be easily inverted.

\subsection{Improving the Rank Histogram Filter}
The RHF described in the preceding section is extremely convenient and quite general, though no proof of its convergence as $N\to\infty$ has yet been provided.
This subsection proposes a different algorithm - an improved RHF (iRHF) - motivated by the form and ease of use of the RHF, but making use of more traditional approximations.

\paragraph{Prior PDF} In the first place, the RHF's approximation of the prior pdf is quite novel.
A more traditional approach to scalar density estimation would use kernel density estimation
\begin{equation}
p(z) = \frac{1}{N}\sum_{i=1}^Nh_i^{-1}K\left(\frac{z-z_i}{h_i}\right)
\end{equation}
where the kernel is $K$ and the bandwidth of the $i^\text{th}$ kernel is $h_i$.
In order to preserve the simplicity and speed of the RHF, it is here proposed to use the most basic kernel: an indicator function (also known as a `top hat' or `boxcar' function)
\begin{equation}
K(s) = \left\{\begin{array}{rl}0&\text{ for }s\le-\frac{1}{2}\text{ or }s\ge\frac{1}{2},\text{ and}\\
1&\text{ for }-\frac{1}{2}<s<\frac{1}{2}.\end{array}\right.
\end{equation}
The bandwidth is estimated using the standard rule-of-thumb for the Gaussian \cite{Silverman98}, with the prefactor modified to fit the top-hat kernel
\begin{equation}
\bar{h} = 3.13\;\text{min}\left\{\hat{\sigma},\frac{\text{IQR}}{1.34}\right\}N^{-1/5}
\end{equation}
where $\hat{\sigma}$ denotes the empirical standard deviation of the prior sample, and IQR denotes the inter-quartile range.
It is possible with this bandwidth for $p(z)$ to have zero probability density over some subintervals; this happens if there is any $i$ for which $z_{k_{i+1}}-z_{k_i}>\bar{h}$.
This would be undesirable, because it would imply that the approximation to the posterior cdf $F_+$ would no longer be invertible.
To avoid this situation, the kernel bandwidth is allowed to vary as follows.
For $i=2,\ldots,N-1$
\begin{eqnarray}
h_i = \frac{1}{2}\max\left\{z_{k_{i+1}}-z_{k_i},z_{k_i}-z_{k_{i-1}},2\bar{h}\right\}
\end{eqnarray}
and
\begin{eqnarray}
h_1 &=& \frac{1}{2}\max\left\{z_{k_2}-z_{k_1},2\bar{h}\right\},\\
h_N&=&\frac{1}{2}\max\left\{z_{k_N}-z_{k_{N-1}},2\bar{h}\right\}.
\end{eqnarray}
This approximation, like that of the RHF, is piecewise-constant on the interior subintervals, though this new approximation has $2N-1$ interior subintervals rather than $N-1$.
The discontinuities occur at $z_m$, $m=1\ldots,2N$, and the locations can be found simply by sorting $z_{k_i}\pm h_i$.
The prior cdf $F_Z$ is piecewise linear, and can be evaluated easily, though not quite so easily as for the RHF.
This approximation of the posterior cdf is used to evaluate $F_Z(z_i)$ in the update $z_i^+=F_+^{-1}(F_Z(z_i))$.
The approximation used here could be updated using, e.g., a triangular or quadratic kernel.
Such an approximation would be piecewise-polynomial and therefore easy to handle numerically, but the resulting differences in performance might be small.

\paragraph{Posterior CDF} The foregoing approximation to the prior pdf has no tails, i.e. the probability of $z$ is zero outside some bounded interval.
This would be problematic whenever the likelihood is large outside the support of $p(z)$, which is not unusual.
The approximation of the prior pdf is therefore updated before using it to generate an approximation of the posterior pdf.
Specifically, the new approximation to the prior is
\begin{equation}
\hat{p}(z) = \frac{1}{P_1}\left(\text{tails} + \frac{1}{N}\sum_{i=1}^Nh_i^{-1}K\left(\frac{z-z_i}{h_i}\right)\right)
\end{equation}
where the `tails' are those of a Gaussian pdf using the empirical mean and variance of the prior sample.
Unlike in the RHF, these tails are not each normalized to have total probability $1/(N+1)$.
The normalization constant $P_1$ is used to force total probability 1, and is available in closed form.
(When the range of $Z$ is not $\mathbb{R}$, standard kernel density estimation techniques like transforms and reflections rather than Gaussian tails can be used to deal with boundaries \cite{Silverman98}.)

With this new approximation of the prior pdf, we simply need a piecewise approximation of the likelihood.
It is here proposed to use a shape-preserving piecewise-cubic interpolant on the interior subintervals (in the numerical experiments this is provided by Matlab's {\tt pchip} interpolator), and the same constant approximation as used by the RHF in the tails.
Constructing this approximation requires evaluating the likelihood at $2N$ points as compared to the RHF's $N$ evaluations of the likelihood, which might lead to cost considerations, depending on the application.
A shape-preserving interpolant is used to avoid producing negative values of the likelihood in between the interpolation points, which would lead to a non-invertible approximation of the posterior cdf.

With these approximations of the prior and likelihood, the posterior pdf is approximated as a piecewise polynomial with Gaussian tails.
The posterior cdf is approximated as
\begin{equation}
F_+(z) \approx \frac{1}{P_2}\int_{-\infty}^z\hat{p}(s)\ell(s)\mathrm{d}s
\end{equation}
where the normalization constant $P_2=\int_\mathbb{R}\hat{p}(z)\ell(z)$d$z$ is available in closed form, as is the value of $F_+(z)$ on any subinterval.
This approximation is an error function in the tails, and piecewise-quartic in the interior.

\paragraph{Inverting the Posterior CDF} The piecewise-quartic posterior cdf of the iRHF is not as easy to invert as the piecewise-quadratic posterior cdf of the RHF.
A good first guess for a rootfinding algorithm would be to linearly interpolate the posterior cdf at the subinterval boundaries, since it is trivial to invert a piecewise-linear monotonic function. 
We propose to take this a step further, by simply using this first guess without further rootfinding. 
(Rootfinding is still carried out for values in the Gaussian tails.)
In our numerical experiments we find that this time-saving approximation has no discernible impact on the results.

\subsection{Comparing the RHF and iRHF in simple examples}
\begin{figure}
\centering
\includegraphics[width=.45\textwidth]{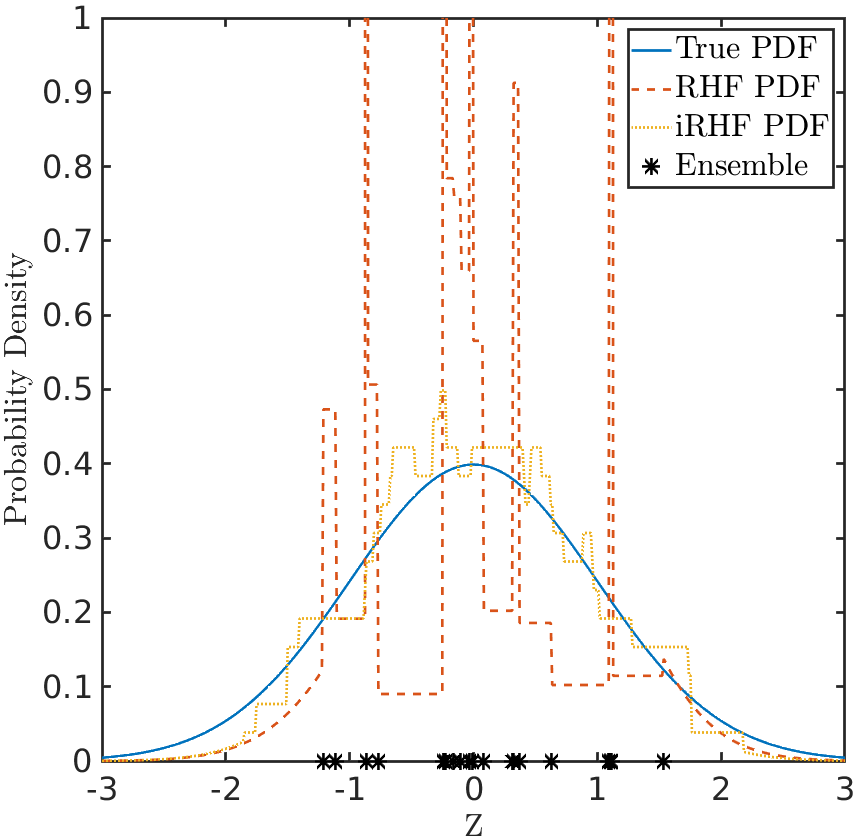}
\caption{The standard normal pdf, together with the approximations produced by the RHF and iRHF. The ensemble used to construct the RHF and iRHF approximations is marked along the horizontal axis.}
\label{fig:PP}
\end{figure}

This section compares the RHF and iRHF for Gaussian problems.
The RHF and iRHF approximations to a standard normal pdf are first compared by drawing an ensemble of $N=20$ samples from $\mathcal{N}(0,1)$.
These same samples are used to construct the RHF and iRHF approximations to the prior pdf, which are plotted in Figure \ref{fig:PP}, together with the standard normal pdf.
Although the RHF and iRHF both use crude piecewise-constant approximations, the iRHF approximation is clearly superior.
Because RHF assigns equal probability to regions between ensemble members, whenever two ensemble members are close the pdf has a large spike. 
For example, there is a large spike in the RHF approximation to the PDF near $z=-1$ in Figure \ref{fig:PP} that extends to a height of approximately 39.

Next consider a scalar problem where the prior on $Z$ is Gaussian with mean $\mu_z$ and variance $\sigma_z^2$, and the observation $y$ is a draw from $Y = Z + \eps$ with $\eps\sim\mathcal{N}(0,\gamma^2)$.
Without loss of generality, let $\mu_z=0$ and $\sigma_z^2=1$.
In this case, the map from samples of the prior to samples of the posterior is affine and is given by the formula
\begin{equation}
z_i^+ = \frac{y}{\gamma ^2+1} + \frac{\gamma}{\sqrt{1+\gamma^2}}z_i = F_+^{-1}(F_Z(z_i)).
\end{equation}

\begin{figure*}
\centering
\includegraphics[width=.8\textwidth]{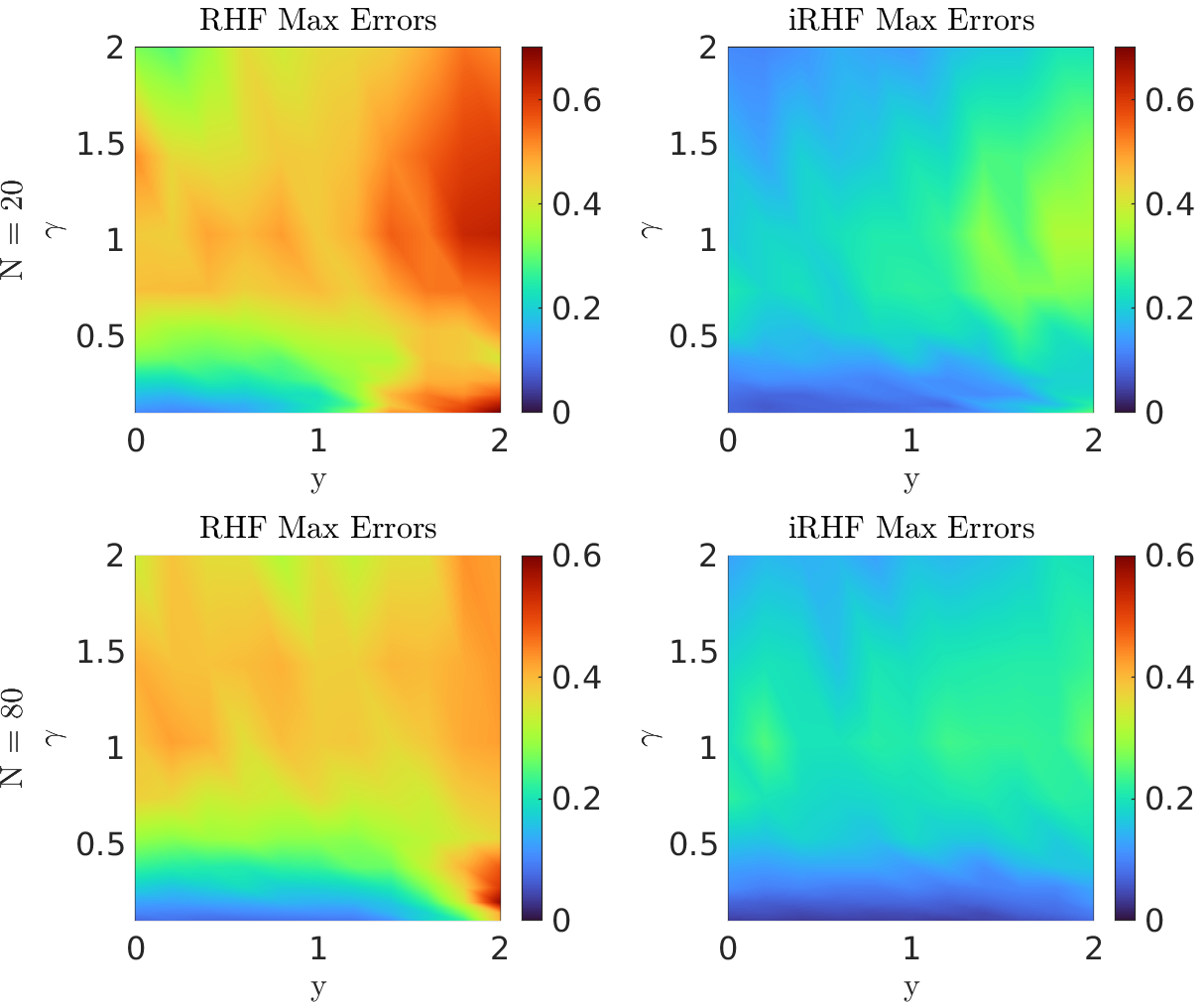}
\caption{Maximum absolute error between the correct value of $z_i^+$ and the values reported by the RHF (left) and iRHF (right) filters, reported as a median over 100 independent trials, as a function of $y$ (the observation value) and $\gamma$ (the observation error standard deviation), for both $N=20$ (upper) and $N=80$ (lower).}
\label{fig:1}
\end{figure*}

To explore the performance of the RHF and the iRHF in approximating this affine map we perform a suite of experiments.
The observation value $y$ and the observation error $\gamma$ are both ranged from 0 to 2.
For each parameter value we run 100 independent experiments at an ensemble size $N=20$, and another 100 independent experiments at $N=80$.
For each experiment the maximum absolute error in $z_i^+$ is recorded; the median over all 100 trials is shown in Figure \ref{fig:1} as a function of $y$ and $\gamma$ for both filters and for $N=20$ and $80$.
Results using the root mean squared error in $z_i^+$ are qualitatively similar (not shown).

The most basic conclusion is that for these examples the iRHF performs more accurately than RHF, and that performance improves from $N=20$ to $N=80$, though the RHF at $N=80$ is still not as accurate as the iRHF at $N=20$.
Both filters perform well when the observation error $\gamma$ is small and the observation $y$ is within one standard deviation of the prior mean.
Both filters' performance decreases when either $\gamma$ increases or when $y$ is far from the prior mean; both of these correspond to situations where the likelihood has large values in the shoulders or tails of the prior distribution.
This is not surprising; since the RHF and iRHF approximations are based on samples from the prior, the RHF and iRHF filters will both have worse representations of the transformation in regions where there are few samples.
Probably the crudest approximation made in the RHF and iRHF is to treat the likelihood as constant in the tails.
Generally speaking, one expects to know the exact functional form of the likelihood function for every observation assimilated, and this information could be used to improve the approximation.
That said, it is not clear if there is a general strategy that works better than the constant approximation for a wide range of likelihoods, without incurring too much extra cost.

\section{Numerical Experiments: Configuration}
This section describes the configuration of a suite of data assimilation experiments and their results, with the goal of comparing a range of nonlinear extensions of the EnKF.
All code used in the numerical experiments is available at \cite{NEnKF}.

\subsection{Model Configuration}
The dynamical model is the Lorenz-`96 model \cite{Lorenz96}
\begin{equation}
\dd{x_k}{t} = -x_{k-1}(x_{k-2}-x_{k+1}) -x_k+F
\end{equation}
with $F=8$, $k=1,\ldots,40$, and the periodic convention $X_{k\pm40} = X_k$.
A reference solution is initialized from a standard normal distribution and integrated forward by 9 nondimensional Model Time Units (MTUs). 
According to the classical accounting based on comparing the error doubling time for this model to that for midlatitude synoptic atmospheric dynamics 0.2 MTUs correspond to one `day' \cite{LE98}.
Starting at $t=9$ MTU, a reference solution is saved every 0.05 MTUs for a total of 5,500 time points.
Time integration is carried out using Matlab's {\tt ode45}, which internally uses a fourth-order Runge-Kutta method with an adaptive step size.
Although the degree of nonlinearity in a forecast depends on both the initial distribution and the length of the time window, 0.05 MTUs is fairly short for this model - not long enough to generate a strongly non-Gaussian forecast.

\subsection{Observing Systems}
Three observing systems are configured.
The first is direct observations of all model variables 
\begin{equation}
\bm{y} = \bm{x} + \bm{\eps}
\end{equation}
where $\bm{x}$ is the reference solution and $\bm{\eps}\sim\mathcal{N}(\bm{0},\mat{I})$.
This configuration is referred to as `linear observations' and is provided simply for reference, as a means to validate the performance of all methods in a well-studied and nearly-Gaussian problem.

The second observing system takes the form
\begin{equation}
\bm{y} = \frac{1}{1 + \text{exp}\{0.5\times(\bm{x}-2.5)+\bm{\eps}\}}
\end{equation}
where the observation error is again standard normal $\bm{\eps}\sim\mathcal{N}(\bm{0},\mat{I})$.
These observations are bounded between 0 and 1, and the expression $0.5\times(\bm{x}-2.5)$ was chosen so that observations will cover nearly the whole range of values from 0 to 1. 
The error distribution can be considered `logit-normal' in the sense that
\begin{equation}
\text{logit}(\bm{Y}) = \log\left(\frac{\bm{Y}}{1-\bm{Y}}\right)
\end{equation}
is Gaussian for fixed $\bm{x}$.
This configuration is referred to as `logit-normal observations.'

Although the observation operator is clearly nonlinear, the likelihood function is actually Gaussian.
Nevertheless, the EnKF does not `approximate the likelihood as Gaussian' - in fact it makes no use of the likelihood function at all.
Instead, as noted in section \ref{sec:EnKF}, the EnKF approximates the joint distribution of $\bm{X}$ and $\bm{Y}$ as Gaussian, which is far from true for this observation operator.
Whenever $\bm{X}$ is normally distributed, the joint distribution of $\bm{Y}$ and $\bm{X}$ has a Gaussian copula, so GA methods should converge towards the true distribution.

The third observing system takes the form
\begin{equation}
\bm{y} = \text{exp}\{0.5\times|\bm{x}-2.5|+\bm{\eps}\}
\end{equation}
where the observation error is again standard normal $\bm{\eps}\sim\mathcal{N}(\bm{0},\mat{I})$.
The observations are positive, and the error distribution is log-normal in the sense that
\begin{equation}
\log(\bm{Y}) = 0.5\times|\bm{x}-2.5|+\bm{\eps}
\end{equation}
is Gaussian for fixed $\bm{x}$.
The likelihood is bimodal because of the absolute value, and the joint distribution of $\bm{X}$ and $\bm{Y}$ does not have a Gaussian copula when $\bm{X}$ is Gaussian.
The expression $0.5\times|\bm{x}-2.5|$ ensures that bimodality primarily occurs for values of $\bm{x}$ near 2.5, which is close to the time-mean value of $\bm{x}$, so that bimodality occurs frequently.

\subsection{Performance Metrics}
The methods compared are described in the following subsections.
Each method generates an ensemble forecast and an ensemble analysis.
The first performance metric is the root mean squared error (RMSE) in the ensemble mean, defined as
\begin{equation}\label{eqn:RMSE}
\left[\frac{1}{40}\sum_{k=1}^{40}(x_k-\bar{x}_k)^2\right]^{1/2},\;\;\bar{\bm{x}} = \frac{1}{N}\sum_{i=1}^N\bm{x}_i
\end{equation}
where $x_k$ denotes the $k^\text{th}$ element of the true solution vector $\bm{x}$.
The forecast and analysis RMSE are stored at each of the 5,500 assimilation cycles, and the median values over the last 5,000 cycles are reported as a summary.
The forecast and analysis spread are also tracked; the spread is defined as
\begin{equation}
\left[\frac{1}{40}\sum_{k=1}^{40}\text{Var}_i[x_k]\right]^{1/2}
\end{equation}
where Var$_i[x_k]$ denotes the ensemble variance of the variable $x_k$.
Generally speaking, one expects the spread to match the RMSE.
Finally, the continuous ranked probability score (CRPS; \cite{Hersbach00,GR07}) is tracked for each variable $x_k$ at each assimilation cycle.
The CRPS measures the quality of a probabilistic estimate, with 0 being the best possible score (though it bears noting that the true Bayesian posterior does not have a CRPS of 0).

\subsection{Overview of Methods}\label{sec:ListOfMethods}
The methods compared here are 
\begin{itemize}
\item A perturbed-observation EnKF with simultaneous assimilation of all observations.
This method is labeled `EnKF.'
The analysis update is based on the conditional-Gaussian formula (\ref{eqn:GaussCondMean}) rather than the EnKF formulas (\ref{eqn:KF_Mean}) \& (\ref{eqn:KF_Gain}). While this is not the best EnKF for the linear-obs setup, it is crucial for the logit-normal and log-normal configurations where the observing system does not take the traditional form $\bm{y} = \bm{H}(\bm{x})+\bm{\eps}$. Also, in some nonlinear and non-Gaussian cases the perturbed-observation EnKF can be more robust than deterministic versions \cite{LBS10}.
\item Two perturbed-observation GA-EnKF methods using the algorithm from section \ref{sec:GA}, which differ in the way they construct the univariate anamorphosis transformations. One method uses a piecewise-linear anamorphosis transformation and is labeled `GA-PL' while the other uses kernel density estimation and is labeled `GA-KDE.' Further details are given in section \ref{sec:GA_Details}.
\item Two two-step methods: The RHF with linear regression and the iRHF with linear regression; these are labeled `RHF' and `iRHF', respectively. These methods use serial assimilation, and the scalar auxiliary variable $Z$ is always chosen to be equal to $x_k$, where $x_k$ is the variable being observed. This choice is made so that as much non-Gaussianity as possible can be dealt with in the first part of the two-step cycle, leaving as much Gaussianity as possible for the linear regression step.
\end{itemize}
For all of these methods the ensemble is initialized by adding draws from a standard normal distribution to the reference solution.

There is a huge range of nonlinear data assimilation methods that could be compared.
Since the focus here is on nonlinear and non-Gaussian extensions of the EnKF, particle filters and variational methods are not implemented.
Other notable classes of methods that are omitted here include Gaussian mixture methods, e.g. \cite{AA99,BSN03,FK13b,LAH16}, the method of \cite{MCSB14}, and the Marginal Adjustment RHF \cite{Anderson20}.
The GIGG-EnKF of \cite{Bishop16,PB18} is not used because the observations do not have Gamma or Inverse-Gamma distributions.

Polynomial regression was implemented for the second step of the RHF and iRHF filters, but initial tests indicated that it was highly unstable due to the gigantic increments produced whenever the observation lies outside the ensemble spread.
For the specific configurations here, where the relationship between $Z$ and $\bm{X}$ is not strongly nonlinear, regression models more advanced than linear, including the rank regression method of \cite{Anderson19}, are unlikely to lead to significant improvements in performance.

For small-scale problems such as this one, it is valuable to compute, when possible, a consistent approximation of the Bayesian posterior, e.g.~using Markov Chain Monte Carlo (MCMC) \cite{LS12,ILS13}.
A sequential importance resampling (SIR) particle filter with an ensemble size of $N=3.072\times10^5$ with particle rejuvenation was run as an attempt to obtain a consistent approximation of the true posterior.
Even with this extremely large ensemble size the SIR particle filter provided very poor estimates, much worse than the competing methods.

\subsection{GA Method details}\label{sec:GA_Details}
As noted in section \ref{sec:GA}, the correct transformation of a scalar random variable $X$ to a Gaussian random variable $\hat{X}$ is $\hat{X} = \Phi^{-1}\left(F_X(X)\right)$ where $F_X$ is the cdf of $X$ and $\Phi$ is the cdf of the Gaussian variable $\hat{X}$.
(This assumes that $F_X$ is continuous.)
A straightforward approach to approximate this transformation is to use kernel density estimation to obtain an approximation to $F_X$, and then to let $\Phi$ be the cdf of a standard normal.
The method GA-KDE tested here uses Matlab's {\tt ksdensity} to estimate $F_X(X)$ for each element of $\bm{x}$ and $\bm{y}$.
The {\tt ksdensity} method uses a Gaussian kernel with the standard rule-of-thumb bandwidth to approximate the prior pdf.
The cdf is then approximated as a sum of error functions, and the inverse is obtained using a rootfinding algorithm.
This approach would be overly costly for large-scale applications, but is useful here as a benchmark for careful and accurate approximation of both $F_X$ and its inverse.

The cdf must be estimated for each element of $\bm{Y}$, which have bounded support in the logit-normal and log-normal configurations.
The default behavior of {\tt ksdensity} for these situations is to transform the data to unbounded variables before applying the standard kernel density estimation procedure.
For the logit-normal data {\tt ksdensity} transforms the data using the logit function, and for the log-normal data it transforms the data using the log function.

The straightforward approach of the GA-KDE method has not been previously used in the GA literature.
Simon \& Bertino \cite{SB12} use a complicated transformation whose main features are emulated here using a method labeled `GA-PL' where PL stands for piecewise-linear.
Simon \& Bertino use a very similar estimate of $F_X$ to what is used in the RHF.
Letting $x_{k_1}<\ldots<x_{k_N}$ be the sorted prior ensemble, and noting that the transformation from $X$ to $\hat{X}$ only needs to be applied to the prior ensemble $x_i$, simply let
\begin{equation}
F_X(x_{k_i})= \frac{i}{N+1}\;.
\end{equation}
The full transform becomes
\begin{equation}\label{eqn:SB_Forward}
\hat{x}_{k_i} = \Phi^{-1}\left(\frac{i}{N+1}\right)
\end{equation}
where $\Phi$ is the standard normal cdf.
Note that the output of this transformation is independent of the actual values of $x_i$.

In the case of the observations, one has both an ensemble of perturbed observations $y_i$, $i=1,\ldots,N$, and the observation itself $y$.
The above formula is used to transform the ensemble of perturbed observations, and the observation $y$ is transformed to $\hat{y}$ by linearly interpolating the map (\ref{eqn:SB_Forward}).
In the case where $y$ falls outside the range of the ensemble $y_i$ direct linear interpolation is not possible.
Simon \& Bertino use exponential tails to extrapolate, whereas GA-PL simply adds extra points and then linearly interpolates.
Specifically, for linear obs GA-PL adds the points $-10 = \Phi^{-1}\left(F_Y(\bar{y}-10\sigma_y)\right)$ and $10 = \Phi^{-1}\left(F_Y(\bar{y}+10\sigma_y)\right)$ where $\sigma_y$ is the standard deviation of the ensemble of perturbed observations.
For logit-normal observations the values $0$ and $1$ are mapped to $-20$ and $20$, respectively.
For log-normal observations the value $0$ is mapped to $-20$, and the value $\bar{y}+4\sigma_y$ is mapped to $4$.
(Preliminary testing indicated better performance from mapping $\bar{y}+4\sigma_y$ to $4$ than from mapping $\bar{y}+10\sigma_y$ to $10$.)

The foregoing discussion explained that the map from $X$ to $\hat{X}$ (and from $Y$ to $\hat{Y}$) is piecewise-linear for GA-PL.
The inverse map, which is used to transform the analysis ensemble $\hat{X}$ to the analysis ensemble $X$, is easily obtained since it is also piecewise-linear, and is implemented using linear interpolation.

\subsection{Inflation and Localization}
Inflation and localization are crucial for the performance of EnKF-type methods.
For the EnKF, RHF, and iRHF methods inflation is implemented by multiplying the forecast ensemble perturbations by a positive inflation factor $r$ before assimilation
\begin{equation}
\bm{x}_i\leftarrow \bar{\bm{x}} + r\left(\bm{x}_i-\bar{\bm{x}}\right).
\end{equation}
where $\bar{\bm{x}}$ is defined in (\ref{eqn:RMSE}).
For the GA methods inflation is implemented in the transformed space
\begin{equation}
\hat{\bm{x}}_i\leftarrow \hat{\bar{\bm{x}}} + r\left(\hat{\bm{x}}_i-\hat{\bar{\bm{x}}}\right).
\end{equation}

All methods use Schur-product localization.
The localization matrix is populated using an exponential localization function
\begin{equation}
\left(\mathbf{L}\right)_{ij} = \text{exp}\left\{-\frac{1}{2}\left(\frac{d_{ij}}{d}\right)^2\right\}
\end{equation}
where $d_{ij}=\min\{|i-j|,40-|i-j|\}$ and $d$ is the localization radius.
In the EnKF method the ensemble approximations to both $\Cov[\bm{X},\bm{Y}]$ and $\Cov[\bm{Y}]$ are multiplied by $\mat{L}$ elementwise.
In the GA methods the ensemble approximations to both $\Cov[\hat{\bm{X}},\hat{\bm{Y}}]$ and $\Cov[\hat{\bm{Y}}]$ are multiplied by $\mat{L}$ elementwise.
In the RHF and iRHF methods, the analysis increments produced by linear regression are localized based on the distance between $x_k$ and $z$.

The localization matrix $\mat{L}$ is not guaranteed to be positive definite using the distance $d_{ij}$.
This distance was replaced by embedding the 40 Lorenz-`96 variables on a circle in $\mathbb{R}^2$ before computing distances, and tested using the EnKF method in the linear problem.
No significant differences were observed.

For all methods, a range of inflation factors and localization radii are tested.
Inflation factors and localization radii are tested in the range
\begin{eqnarray}
r &=& 1 + \frac{k}{20}, k=0,\ldots,8\\
d &=& 0.5, 2k-1\text{ for }k=1,\ldots,8,\text{ and }\infty
\end{eqnarray}
A wide range was required because the optimal parameter values differed significantly between methods.

\section{Numerical Experiments: Results}
We begin by reporting results on the linear-obs problem, primarily as a baseline to verify that the methods are working as expected.
Results are only reported for an ensemble size of $N=120$.
Some interesting results arise, which are briefly discussed before turning to the main topic of the nonlinear, non-Gaussian problems.

\subsection{Linear Observations}
Table \ref{tab:lin} presents the results for all methods in the problem with linear observations at the highest ensemble size $N=120$.
All methods perform well, with RMS errors well below the observation error level (which is 1), but the RHF and iRHF perform significantly better than the EnKF and GA methods.
This problem is only weakly nonlinear, i.e. the distributions are only weakly non-Gaussian, so the nonlinear extensions to the EnKF are not expected to produce substantial improvements in performance.
This suggests that the improvements seen in the RHF and iRHF methods are primarily the result of better stability to sampling errors, which is supported by the smaller optimal inflation factors and larger optimal inflation radii.
It is worth recalling that the EnKF implemented here is a perturbed-observation version of the EnKF, which can be more susceptible to sampling errors than deterministic EnKFs, which further supports the idea that the difference between the EnKF and RHF is mainly due to the different sensitivity to sampling errors.
The GA-PL and GA-KDE methods have the same perturbed-observation EnKF inside the transformed space, so they are also more susceptible to sampling errors.
Thus, the three perturbed-observation methods perform similarly, and the two RHF methods perform similarly.
The EnKF, GA-PL, and GA-KDE methods have slightly under-dispersed ensembles, while the RHF and iRHF methods have well-spread ensembles.
Since the focus is on non-Gaussian DA, the results of the linear observation setup are not analyzed further.

\begin{table*}
\caption{Results for all methods in the linear observation problem with ensemble size $N=120$. For each method results are given at the parameter values that produce the optimal analysis RMSE. Forecast RMSE is abbreviated F.~RMSE, and Analysis RMSE is abbreviated A.~RMSE, etc.}
\label{tab:lin}
\begin{tabular}{lllcccccc}
\hline\noalign{\smallskip}
Method & Loc.~Radius & Inflation & F.~RMSE & F.~Spread & F.~CRPS & A.~RMSE & A.~Spread & A.~CRPS \\
\noalign{\smallskip}\hline\noalign{\smallskip}
EnKF 		& 3 	& 1.05 & 0.28 & 0.25 & 0.11 & 0.26 & 0.23 & 0.10  \\
GA-PL 		& 3 	& 1.05 & 0.29 & 0.26 & 0.11 & 0.26 & 0.24 & 0.10  \\
GA-KDE 	& 5  	& 1.10 & 0.30 & 0.26 & 0.11 & 0.28 & 0.23 & 0.10  \\
RHF 		& 15 	& 1 	& 0.19 & 0.18 & 0.07 & 0.17 & 0.17 & 0.06  \\
iRHF & 	$\infty$ & 1	& 0.19 & 0.20 & 0.07 & 0.17 & 0.18 & 0.06  \\
\noalign{\smallskip}\hline
\end{tabular}
\end{table*}

\subsection{Logit-normal Observations}
Table \ref{tab:logit} presents the results for all methods in the problem with logit-normal observations at ensemble size $N=120$.
The EnKF does quite well despite the nonlinear, non-Gaussian observations, with small RMS errors and a slightly under-dispersed ensemble. 
Adding Gaussian Anamorphosis to the EnKF via the piecewise-linear transform in the GA-PL method degrades the analysis RMSE by 11\%, while the GA-KDE method leads to a 5\% improvement in analysis RMSE.

As in the linear observation problem, the RHF and iRHF methods are significantly better than the EnKF and GA methods, and they also require less localization and less inflation.
They both produce well-balanced ensembles.
\begin{table*}
\caption{Results for all methods in the logit-normal observation problem with ensemble size $N=120$. For each method results are given at the parameter values that produce the optimal analysis RMSE. Forecast RMSE is abbreviated F.~RMSE, and Analysis RMSE is abbreviated A.~RMSE, etc.}
\label{tab:logit}
\begin{tabular}{lllcccccc}
\hline\noalign{\smallskip}
Method & Loc.~Radius & Inflation & F.~RMSE & F.~Spread & F.~CRPS & A.~RMSE & A.~Spread & A.~CRPS \\
\noalign{\smallskip}\hline\noalign{\smallskip}
EnKF 		& 3 	& 1.05 & 0.60 & 0.55 & 0.23 & 0.55 & 0.50 & 0.21  \\
GA-PL 		& 3 	& 1.05 & 0.67 & 0.56 & 0.25 & 0.61 & 0.51 & 0.23  \\
GA-KDE 	& 3  	& 1.05 & 0.57 & 0.49 & 0.22 & 0.52 & 0.45 & 0.20  \\
RHF 		& 9 	& 1 	& 0.42 & 0.40 & 0.16 & 0.39 & 0.36 & 0.14  \\
iRHF 		&15  	& 1		& 0.41 & 0.43 & 0.15 & 0.38 & 0.39 & 0.14  \\
\noalign{\smallskip}\hline
\end{tabular}
\end{table*}

\begin{figure}
\centering
\includegraphics[width=.45\textwidth]{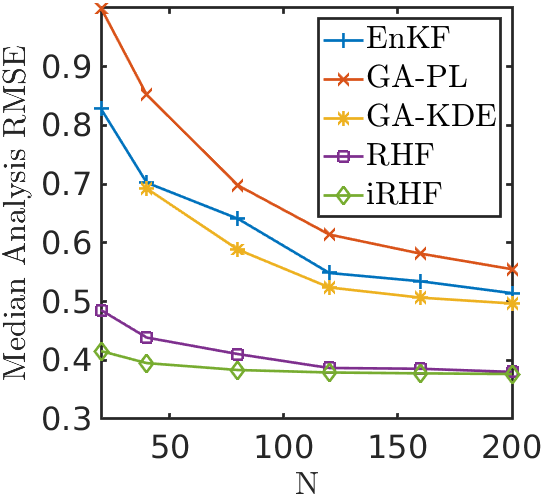}
\caption{Analysis RMSE in the problem with logit-normal observations as a function of ensemble size $N$ for all methods. The GA-KDE method was unstable at $N=20$ and failed to produce results at any of the parameter values tested. The methods RHF and iRHF appear converged at $N=120$, so results were not computed for $N>120$.}
\label{fig:2}
\end{figure}

The patterns evident at $N=120$ persist as the ensemble size is varied.
Figure \ref{fig:2} shows the time-median analysis RMSE for all methods as a function of ensemble size $N$.
Of particular note is the fact that the RHF and iRHF methods at $N=20$ still perform significantly better than the other methods at $N$ up to 200.
Also, the difference between RHF and iRHF increases as the ensemble size is lowered, with RHF having 17\% worse analysis RMSE than iRHF at $N=20$.
Finally, the RHF and iRHF results appear to have converged by $N=120$, whereas the other three methods require $N>120$ to converge, which further underscores that RHF and iRHF are less sensitive to sampling errors.

\subsection{Lognormal Observations}
Table \ref{tab:log} presents the results for all methods in the problem with log-normal observations at the highest ensemble size $N=120$.
The EnKF is highly unstable with only a few parameter values avoiding blowup and returning results.
In those cases the spread is close to machine precision, and the RMS errors are enormous: The filter completely ignores the observations.
Adding Gaussian Anamorphosis to the EnKF makes a huge difference here.
The GA-KDE method does about 13\% better than the GA-PL method, though both perform reasonably well.
The GA-KDE method has a slightly over-dispersed ensemble while the GA-PL method has a significantly under-dispersed ensemble, though these differences do not end up leading to large differences in the CRPS.
It is possible that the optimal inflation parameter for both methods lies between 1.05 and 1.10.

As in the linear and logit-normal observation problems, the RHF and iRHF methods are significantly better than the EnKF and GA methods, and they also require less localization and less inflation.
They both produce well-balanced ensembles.
\begin{table*}
\caption{Results for all methods in the log-normal observation problem with ensemble size $N=120$. For each method results are given at the parameter values that produce the optimal analysis RMSE. Forecast RMSE is abbreviated F.~RMSE, and Analysis RMSE is abbreviated A.~RMSE, etc.}
\label{tab:log} 
\begin{tabular}{lllcccccc}
\hline\noalign{\smallskip}
Method & Loc.~Radius & Inflation & F.~RMSE & F.~Spread & F.~CRPS & A.~RMSE & A.~Spread & A.~CRPS \\
\noalign{\smallskip}\hline\noalign{\smallskip}
EnKF 		& 7 	& 1 	& 5.20 & 0      & 3.64 & 5.20 & 0      & 3.64  \\
GA-PL 		& 3 	& 1.05 & 0.91 & 0.64 & 0.33 & 0.83 & 0.59 & 0.31  \\
GA-KDE 	& 3  	& 1.10 & 0.78 & 0.87 & 0.31 & 0.72 & 0.80 & 0.29  \\
RHF 		&11 	& 1 	& 0.44 & 0.42 & 0.16 & 0.41 & 0.39 & 0.15  \\
iRHF 		&11  	& 1		& 0.45 & 0.47 & 0.17 & 0.41 & 0.43 & 0.15  \\
\noalign{\smallskip}\hline
\end{tabular}
\end{table*}

\begin{figure}
\centering
\includegraphics[width=.45\textwidth]{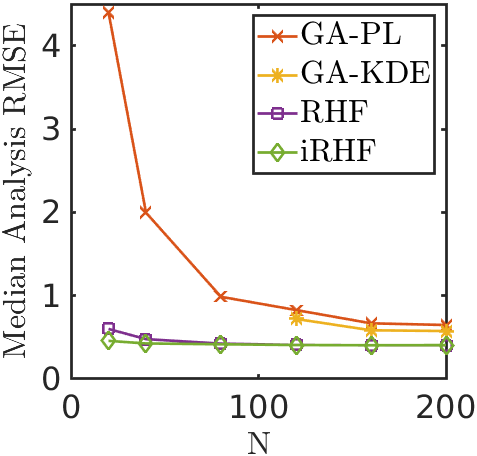}
\caption{Analysis RMSE in the problem with log-normal observations as a function of ensemble size $N$ for all methods except EnKF, whose RMSE is off the chart. The GA-KDE method was unstable and failed to produce results at any of the parameter values tested for $N<120$. RHF and iRHF appear converged at $N=120$, so results were not computed for $N>120$. }
\label{fig:3}
\end{figure}

As in the logit-normal problem, the patterns evident at $N=120$ persist as the ensemble size is varied.
Figure \ref{fig:3} shows the analysis RMSE for all methods as a function of ensemble size $N$.
Once again the RHF and iRHF methods at $N=20$ perform significantly better than the other methods at $N$ up to 200: RHF at $N=20$ is comparable to GA-PL, and GA-KDE at $N=200$, but iRHF at $N=20$ remains significantly better than those methods at $N=200$.
The difference between RHF and iRHF increases as the ensemble size is lowered, with RHF having 33\% worse analysis RMSE than iRHF at $N=20$.
Finally, the RHF and iRHF results appear to have converged by $N=80$, whereas the other three methods require $N>120$ to converge, which again underscores that RHF and iRHF are less sensitive to sampling errors than EnKF and the GA methods.

GA-KDE was particularly unstable on this problem at low ensemble sizes, and diverged to machine infinity before completing 5,500 assimilation cycles for all parameters tested and all $N<120$.
By contrast, GA-PL remained stable and produced better results than EnKF, despite having very poor performance at small ensemble sizes.

\section{Conclusions}
This paper reviews two extensions of the EnKF to nonlinear and non-Gaussian problems.
The first class of methods, Gaussian anamorphosis (GA) methods, use univariate transformations to make the univariate marginal distributions of the state and observation vectors Gaussian before applying an EnKF, followed by the inverse transformation back to the original variables.
GA-EnKF methods should presumably converge in the large-ensemble limit for problems with a Gaussian copula.
The second class of methods is based on a two-step framework introduced by Anderson \cite{Anderson03}; the first step is a scalar Bayesian update and the second step is a regression problem.
The equivalence of the two-step process to the Bayesian update was previously known for Gaussian problems; this review clarifies the connection of the two-step process to the original Bayesian problem with no assumption of Gaussianity.
The new perspective on the two-step process enables a class of EnKF-like data assimilation algorithms that can reasonably be expected to converge in the large-ensemble limit for a wide class of non-Gaussian problems.
No analysis of that limit is performed in this paper, which is focused on understanding the methods, clarifying the situations when they are expected to work, and providing examples showing their success in nonlinear, non-Gaussian problems.

Two specific GA-EnKF methods are described and implemented, which differ in the empirical approximations used for their univariate transforms.
GA-PL uses a piecewise-linear transformation modeled on the method from \cite{SB12}, while GA-KDE uses standard kernel density estimation tools to construct the transformations.
Both of the GA-EnKF methods use perturbed-observation EnKFs, and a matching perturbed-observation EnKF is implemented without the transforms for comparison.
Although deterministic EnKFs are sometimes more accurate than perturbed-observation EnKFs, the latter are sometimes more well-behaved in nonlinear and non-Gaussian problems \cite{LBS10}.
In addition to the RHF version of the two-step method from \cite{Anderson10}, a new first step is developed, called iRHF for `improved' RHF.
In the experiments reported here, both the RHF and iRHF methods use simple linear regression for the second step of the two-step process.
Three suites of numerical experiments are described using the Lorenz-`96 model \cite{Lorenz96}: one using linear observations with Gaussian errors, one using logit-normal observations, and one using log-normal observations with a bimodal likelihood.
The logit-normal observation configuration has a Gaussian copula, so the GA-EnKF methods should perform well, while the log-normal observation configuration does not have a Gaussian copula.

In the linear observation experiments the EnKF, GA-PL, and GA-KDE methods perform similarly.
The RHF and iRHF methods perform similarly to each other, and are both more accurate than the other three methods.
This indicates that RHF and iRHF are less sensitive to sampling errors than the EnKF and GA-EnKF methods, though the differences could also be associated with some weak non-Gaussianity from the nonlinear forecast.

In the logit-normal observation experiments, EnKF does quite well.
GA-PL is actually worse than EnKF, though this is evidently because of the low quality of the transformations since GA-KDE performs better than EnKF.
The RHF and iRHF methods are very similar to each other, and are both significantly more accurate than the other three methods.

In the log-normal observation experiment the EnKF is completely unable to produce a meaningful result; it either diverges to machine infinity, or ignores the observations completely.
The GA-PL and GA-KDE methods do quite well, with GA-KDE performing slightly better than GA-PL due to the more-accurate transformations.
This is an example of a problem where GA methods should not converge to the correct posterior, but where they still lead to significant gains over the EnKF.
By contrast, the RHF and iRHF methods perform similarly and far more accurately than the GA-EnKF methods.

The foregoing discussion notes that RHF and iRHF perform similarly to each other in all tests \emph{at large ensemble sizes}.
For small ensembles on the order of $N=20$ iRHF is significantly more accurate than RHF.
This is because the approximations used in the RHF and iRHF are both accurate for large ensemble sizes, but the iRHF uses approximations that remain accurate at small ensemble sizes.
The ensemble size where differences between RHF and iRHF arise depends on the details of the problem at hand.
The iRHF results at $N=20$ were better than the EnKf and GA-EnKF results at $N=200$ in the experiments with logit-normal and log-normal observations.

Future work in this area could be directed towards different regression models for the second step of the two-step process, and towards rigorously proving convergence of GA and two-step methods for classes of non-Gaussian distributions.
Some preliminary results were computed using cubic polynomial regression for the second step of the RHF and iRHF filters.
These were uniformly more unstable than linear regression due to the large increments produced by a polynomial when extrapolating outside the range of the data, which happens occasionally during a long string of DA cycles.
The simple linear regression model used here was quite accurate in all experiments, so the development of more advanced regression models is left to a future work.
A simple example where regression methods more advanced than simple linear models would be needed is when one or more of the state variables has a bounded domain, e.g. where concentrations must remain positive.

\section*{Acknowledgments}
The author is grateful to J.~L.~Anderson for discussions on the two-step framework and the RHF, and to M.~El Gharamti for discussions of Gaussian anamorphosis methods.
Two anonymous reviewers offered suggestions for clarification and improvement of the presentation.
This work utilized resources from the University of Colorado Boulder Research Computing Group, which is supported by the National Science Foundation (awards ACI-1532235 and ACI-1532236), the University of Colorado Boulder, and Colorado State University.

\appendix
\section*{Appendix: Copulas}
This appendix provides a brief, incomplete background on copulas.
For further details, see \cite{Nelson06,SF08}.
Sklar's theorem \cite{Sklar59} relates the joint cdf $H(x_1,\ldots,x_d)$ of a vector of random variables to the marginal cdfs $F_k(x_k)$ and a `copula' $C$:
\begin{eqnarray}\notag
P[X_1<x_1,\ldots,X_d<x_d] &=& H(x_1,\ldots,x_d) \\
&=& C(F_1(x_1),\ldots,F_d(x_d)).
\end{eqnarray}
The copula $C$ is thus a function from the $d$-dimensional hypercube to $[0,1]$; it is uniquely defined when the marginal cdfs are continuous.
The copula encodes the dependence relationships between the variables, in the sense that strictly-increasing transformations of the random variables $X_i$ do not change the copula \cite[Theorem 2.3.4]{Nelson06}.
To give an example relevant to the discussion of GA methods, let $X_1,\ldots,X_d$ be jointly Gaussian random variables with zero mean, unit variance, and covariance matrix $\mat{C}$.
Let $G_1,\ldots,G_d$ be arbitrary invertible cdfs, and suppose that we apply a probability integral transform to each $X_k$, resulting in the new set of random variables $\hat{X}_k = G_k^{-1}(\Phi(X_k))$, where $\Phi$ is the standard normal cdf.
This new set of random variables $\hat{X}_1,\ldots,\hat{X}_d$ is clearly not jointly normal (the marginal pdf of $\hat{X}_k$ is $G_k'(x)$, which is not Gaussian), but the dependence structure between the variables is completely described by a Gaussian copula.
There is a huge menagerie of families of copulas describing different kinds of dependence structure between variables.
By transforming all the marginal distributions to Gaussian distributions and then applying the EnKF, which assumes that the joint distribution is Gaussian, GA methods are evidently assuming that the distribution of the original variables has a Gaussian copula.

\end{document}